\documentclass[reprint, amsmath, amssymb, aps, nofootinbib, superscriptaddress]{revtex4-2}
\usepackage{graphicx}
\usepackage{hyperref}
\usepackage{dcolumn}
\usepackage{bm}
\usepackage{color}
\usepackage{amsmath}
\usepackage[x11names]{xcolor}
\usepackage{orcidlink}

\providecommand{\prd}{Phys. Rev. D}

\providecommand{\apj}{Astrophys. J.}
\providecommand{\mnras}{Mon. Not. R. Astron. Soc.}
\providecommand{\aap}{Astron. Astrophys.}
\providecommand{\jcap}{JCAP}

\begin{document}
\title{Splitting the action: A fully nonlinear fluid method for clustering dark energy}
\author{Yunhao Gao~\orcidlink{0009-0000-6958-1139}}\email{gaoyh@nao.cas.cn}
\affiliation{National Astronomical Observatories, Chinese Academy of Sciences, Beijing, 100101, China}
\affiliation{School of Astronomy and Space Sciences, University of Chinese Academy of Sciences, Beijing, 100049, China}
\affiliation{Institute for Computational Cosmology, Department of Physics, Durham University, Durham DH1 3LE, UK}

\author{Baojiu Li~\orcidlink{0000-0002-1098-9188}}\email{baojiu.li@durham.ac.uk}
\affiliation{Institute for Computational Cosmology, Department of Physics, Durham University, Durham DH1 3LE, UK}
\affiliation{National Astronomical Observatories, Chinese Academy of Sciences, Beijing, 100101, China}

\author{Jie Wang~\orcidlink{0000-0002-9937-2351}}\email{jie.wang@nao.cas.cn}
\affiliation{National Astronomical Observatories, Chinese Academy of Sciences, Beijing, 100101, China}
\affiliation{School of Astronomy and Space Sciences, University of Chinese Academy of Sciences, Beijing, 100049, China}
\affiliation{\mbox{Institute for Frontiers in Astronomy and Astrophysics, Beijing Normal University, Beijing, 102206, China}}

\author{Carlton M. Baugh~\orcidlink{0000-0002-9935-9755}}\email{c.m.baugh@durham.ac.uk}
\affiliation{Institute for Computational Cosmology, Department of Physics, Durham University, Durham DH1 3LE, UK}

\begin{abstract}
    While the background evolution of dark energy is relatively straightforward to model, its perturbations, especially on nonlinear scales, generally require significantly more sophisticated treatment. Existing simulation methods for clustering dark energy --- a large class of interesting theories which are until now not well studied --- usually rely on grid-based solvers and perturbatively derived equations, whose validity in the nonlinear regime remains unknown. In this paper, we introduce a novel fluid method based on splitting the dark energy action into two parts: a kinetic part 
    and a potential part. The resulting fluid equations are fully nonlinear and not reliant on simplifying approximations, making them suitable to describe the small-scale evolution of clustering dark energy theories, such as $k$-essence models with small sound speeds. We demonstrate the numerical implementation of this framework through spherical-collapse simulations of two representative $k$-essence models, evolving dark matter and dark energy simultaneously. The simulations are able to follow the development of dark-energy density perturbations throughout halo collapse, to $\mathcal{O}(\textrm{1}-10)$. By investigating their dependence on the dark-energy sound speed and dark-matter halo mass, we illustrate how dark-energy clustering responds to variations in the competing effects of pressure support and gravitational pull. These results demonstrate the applicability of this new framework to study nonlinear dark-energy clustering across a range of physical conditions, providing the foundation for its future extension to fully three-dimensional cosmological simulations through particle-based algorithms such as smooth particle hydrodynamics (SPH). 
\end{abstract}

\maketitle

\section{Introduction}
The accelerated expansion of the Universe is one of the central open questions in modern cosmology. Within the standard cosmological framework, one of the explanations of this phenomenon invokes dark energy (DE), a component with negative pressure that dominates the present-day energy density of the Universe. While the standard $\Lambda$CDM model, which explains the DE component as a cosmological constant, is the most broadly accepted and adopted in research over a long period of time, recent measurements from DESI have renewed interest in dynamical dark-energy scenarios \cite{DESI01, DESI02, DESI03}: when combined with other cosmological probes, the DESI data imply a preference for an evolving dark-energy equation of state (EoS) over a cosmological constant. These developments motivate a broader exploration of dark-energy models beyond $\Lambda$CDM and, in particular, of their potentially observable signatures in both the background expansion and the growth of cosmic structures.

For the wide variety of dark energy models, particularly those described by scalar fields, the evolution of the dark energy component is generally governed by complicated partial differential equations: the equations of motion (EoM) of the DE theories. Their properties are strongly related to the sound speed, $c_s$, of the DE fluid, which describes the propagation speed of perturbations and determines the radius of the sound horizon $c_s / H$ inside which pressure support suppresses the perturbation growth of dark energy \cite{cs2_and_sound_horizon}. For simple quintessence models \cite{quintessence_theory1, quintessence_theory2, quintessence_theory3}, $c_s$ equals the speed of light, so that the sound horizon is of order the Hubble horizon and dark-energy perturbations are negligible on most sub-horizon scales relevant to structure formation. Their cosmological impact can therefore be described, to a very good approximation, through the homogeneous background evolution alone. In these models, the nonlinear evolution of large-scale structure and dark-matter halos can be studied using conventional $N$-body simulations, with dark energy entering primarily through the modified background expansion \cite{Background_sim01, Background_sim02, Background_sim03, Background_sim04, Background_sim05}.

For models with a slightly lower sound speed, e.g. $c_s^2\sim10^{-2}$--$10^{-3}$, dark-energy perturbations can become non-negligible on sufficiently large scales, while still remaining within the linear-perturbation regime over all scales of interest. Their evolution can therefore be followed by solving the linearized EoM, for example with properly extended Boltzmann solvers such as CAMB \cite{CAMB01, CAMB02} or CLASS \cite{CLASS01, CLASS02}, and subsequently incorporating their gravitational effects into simulations of the dark-matter distribution (some examples are \cite{Linear_sim01, Linear_sim02, Linear_sim03}). Extensions designed for more general parameterizations of dark-energy and modified-gravity theories, such as \texttt{EFTCAMB} \cite{EFTCAMB1, EFTCAMB2, EFTCAMB3} and \texttt{hi\_class} \cite{hiclass1, hiclass2}, provide analogous linear treatments within broader theoretical frameworks.

However, solving the equations of motion of DE analytically and propagating their effects to cosmological observables theoretically becomes almost impossible once large spatial perturbations and nonlinearities are taken into account. This is especially relevant for clustering dark-energy models with a very small $c_s^2$, in which dark-energy perturbations can become significant on small scales and nonlinear effects may no longer be negligible. Perturbation theory has been employed to investigate the nonlinear effects of clustering dark energy \cite{Perturbation_theory01, Perturbation_theory02, Perturbation_theory03}, but its applicability is intrinsically limited to sufficiently large, linear or quasi-linear, scales. As the system enters the deeply nonlinear regime, the perturbative expansion ceases to be well controlled, higher-order corrections become increasingly important, and non-perturbative phenomena cannot be captured reliably. Numerical simulations therefore become an essential tool for following the evolution of such models in this regime. 

A number of previous studies have investigated the evolution of clustering dark-energy theories numerically, using a variety of approaches. Among the simplest are spherical-collapse models \cite{SC01, SC02, SC03, SC04, SC05, SC06} (also see \cite{CDE_review} for spherical collapse as well as a broader review of clustering dark energy), which focus on nonlinear evolution on halo scales and examine how dark-energy perturbations modify the formation of dark-matter halos, in particular quantities such as the critical density threshold for collapse. However, these studies generally rely on simplifying assumptions about the dark-energy dynamics, such as restricting the sound speed to particular limiting cases or adopting quasi-static, linearized, or phenomenological approximations.

For three-dimensional cosmological simulations in which clustering dark energy is evolved as an independent component, two numerical codes have been developed to date. The first is $k$-evolution \cite{k-evolution01} (and its extension to KGB models \cite{k-evolution02}), which depends on the effective field theory of dark energy (EFTofDE) and treats the density perturbation through the Stuckelberg scalar field. This framework has also been employed in some subsequent studies of nonlinear DE dynamics \cite{k-evolution03, k-evolution04}. The second one, \texttt{Nefertiti}, was developed in \cite{FVM01, FVM02}. This code treats the DE component as a fluid with the equation of state $P=w\bar{\rho}c^2+\bar{c}_s^2\delta\rho$ (Eq.(1) in the \cite{FVM02}), and computes the evolution of this fluid through the finite volume method (FVM). Despite their respective advantages, similar limitations to the previously-mentioned spherical collapse models are also present in these studies: both approaches rely on phenomenological parameterizations of the dark energy, which do not directly associate the simulation with a specific underlying DE theory, and both formulations involve expansions in physical quantities assumed to remain perturbatively small: in $k$-evolution, the evolution equations are obtained by expanding in the Stuckelberg scalar perturbation, whereas the equation of state adopted in \texttt{Nefertiti} corresponds to an expansion of the pressure around the homogeneous background. As discussed above, the validity of such expansions becomes less well controlled in the deeply nonlinear regime, where the relevant perturbations may no longer remain small in terms of these expansion quantities. This may have implications for the accuracy or range of applicability of these algorithms on strongly nonlinear scales.

In this paper, we present a detailed formulation and numerical implementation of a new fluid framework for clustering dark energy, derived directly from the underlying scalar-field theory. This framework is based on an action-splitting method, through which we separate the DE action into one kinetic part and one potential part, and we only simulate the kinetic density $\rho_k$ whose EoS is exactly $P_k=\rho_k c^2$ at all scales. This avoids the complicated relation between the pressure and the density when the DE density contrast becomes $\mathcal{O}(1)$. The resulting fluid equations are connected with specific DE theories, instead of parameterizations, and they need no estimations or small-quantity expansions of the fluid quantities. As a first application of the framework, we perform spherical-collapse simulations of two representative $k$-essence models, investigating the nonlinear evolution of dark-energy clustering and its dependence on the sound speed and dark-matter halo mass.

The structure of this paper is as follows. In Section~\ref{sec: equation derivation}, we introduce the idea of action splitting and the derivation of our fluid equations through quintessence, and then generalize them to more complicated $k$-essence models. In Section~\ref{sec: SC}, we apply the fluid framework to spherical-collapse simulations of two $k$-essence models, presenting the evolution of their density profiles and examining the dependence of DE clustering on different physical conditions. Finally, we discuss the results in this paper and conclude in Section~\ref{sec: discussion and conclusion}. 

We leave the derivation of the spherical collapse equations in Appendix~\ref{app: fluid and poisson}, and discuss the subtleties involved in deciding which small quantities (such as the gravitational potentials and peculiar velocities) should be retained or neglected in the equations in the same section. To verify the reliability of our numerical implementation, we also perform a series of convergence tests, whose results are exhibited in Appendix~\ref{app: convergence}. 

\section{Splitting of the DE action and the fluid equations}\label{sec: equation derivation}

To simulate the dark energy using the fluid method, we assume it can be described by a fluid, whose energy-momentum (EM) tensor can be written in the form of
\begin{equation}\label{EM: ideal fluid}
    T^{ab}=(\rho+P)u^au^b+Pg^{ab},
\end{equation}
where $\rho$ is the energy density, $P$ is the pressure, and $u^a$ is the 4-velocity of the fluid. Using the conservation equation $\nabla_aT^{ab}=0$, we can derive the relativistic continuity and Euler equations:
\begin{equation}\label{normal fluid equations}
\begin{aligned}
    u^a\nabla_a\rho+\left(\rho+P\right)\nabla_au^a&=0,\\
    \left(\rho+P\right)u^a\nabla_au^b&=-(g^{ab}+u^au^b)\nabla_aP.
\end{aligned}
\end{equation}
In these equations, we have \textit{five} unknown functions (density, pressure, and 3 degrees of freedom in the 4-velocity, where the normalisation relation $u^au_a=-1$ eliminates 1 degree of freedom), but we only have \textit{four} equations (1 continuity equation and 3 degrees of freedom in the Euler equation, where $u_bu^a\nabla_au^b=0$ eliminates one component of the 4-vector Euler equation). This is not surprising, since these equations do not specify the EoS of the fluid, which introduces the relation between the pressure and the density and is determined by the microscopic physics of the scalar field. To close the system, we'll make use of the EoS.

On the background, the EoS is simply described by a time-dependent factor, $w(t)$, and the relation between the DE pressure and density is
\begin{equation}
    P_{\text{DE}} = w\rho_{\text{DE}}.
\end{equation}
On linear perturbation scales, where density perturbations are small, the EoS (in the DE rest frame) can be written as:
\begin{equation}
    \delta P_{\text{DE, rest}} = \bar{c}_s^2\delta\rho_{\text{DE, rest}}.
\end{equation}
Here, the background sound speed squared, $\bar{c}_s^2$, determines the propagation speed of density perturbations in the DE fluid. These relations can help us fully describe the background and linear evolutions of dark energy. However, it is difficult to extend this kind of EoS (where the pressure can be written as a function of the density, and possibly also of other fluid variables, such as the velocity) to non-linear scales, where the density perturbation of DE, $\delta_{\text{DE}}$, becomes $\mathcal{O}(1)$, and the effects of the clustering dark energy play a significant role. This poses the main challenge in implementing the previous fluid equations in this regime.

In this section, we illustrate how to bypass the difficulties stated above by splitting the DE action into a kinetic part and a potential part and, in the end, obtain a set of fully self-consistent and accurate fluid equations that can be used on non-linear scales. We will clarify the idea of this method and the relevant derivations through quintessence, which is the simplest dark energy scenario, exhibit how to rewrite the continuity and Euler equations in this new framework, and generalize it to more complicated  $k$-essence models. For readers who are not interested in detailed derivations, the resultant fluid equations can be found in Eqs.~\eqref{continuity2: quintessence}-\eqref{phi2: quintessence} and Eqs.~\eqref{continuity2: k-essence}-\eqref{phi2: k-essence}.

As for the conventions used in this paper: we adopt the metric signature $(-, +, +, +)$, and define
\begin{equation}
\begin{aligned}
    X &= g^{ab}\partial_a\phi\partial_b\phi.
\end{aligned}
\end{equation}
Unless otherwise stated, we always set the speed of light, $c$, to $c=1$.

\subsection{Quintessence}
\label{subsection: quintessence}

The action of the quintessence model is specified as
\begin{equation}\label{action: quintessence}
    S=\int \text{d}^4x\sqrt{-g}\left[-\frac{1}{2}g^{ab} \partial_a\phi\partial_b\phi-V(\phi)\right],
\end{equation}
where $g$ is the determinant of the metric $g_{ab}$, and $V(\phi)$ is a function depend on the scalar field $\phi$. This action consists of two terms: the first term, which includes the derivative of the scalar field, is the \textit{kinetic part}, with a Lagrangian density of
\begin{equation}
    \mathcal{L}_{(k)}=-\frac{1}{2}g^{ab} \partial_a\phi\partial_b\phi,
\end{equation}
which is denoted by a lower index $k$. The second term, $V(\phi)$, is the \textit{potential}, whose Lagrangian density is
\begin{equation}
    \mathcal{L}_{(V)}=-V(\phi),
\end{equation}
which is denoted by a lower index $V$. The corresponding EM tensor of these two parts can be computed through the relation
\begin{equation}
    T_{(i)ab}=-\frac{2}{\sqrt{-g}}\frac{\delta\left(\sqrt{-g}\mathcal{L}_{(i)}\right)}{\delta g^{ab}},
\end{equation}
with $i = k$ or $V$. The result is
\begin{equation}\label{EM: quintessence}
\begin{aligned}
    T_{(k)ab}&=\left[-\frac{1}{2}g^{\mu\nu} \partial_\mu\phi\partial_\nu\phi\right]g_{ab} +\partial_a\phi\partial_b\phi,\\
    T_{(V)ab}&=-V(\phi)g_{ab},
\end{aligned}
\end{equation}
with
\begin{equation}\label{EM relation}
    T_{ab}=T_{(k)ab} + T_{(V)ab},
\end{equation}
where $T_{ab}$ is the total EM tensor, which is derived directly by varying the action, Eq.~\eqref{action: quintessence}.

The next step is converting the EM tensor of (only) the kinetic part into the fluid description similar to Eq.~\eqref{EM: ideal fluid}. To achieve this, we define the 4-velocity of the DE fluid as
\begin{equation}\label{4-vel 1}
    u^a=-\frac{\nabla^a\phi}{\sqrt{-X}},
\end{equation}
which automatically satisfies the constraints $g_{ab}u^au^b=-1$ and $u^0>0$ when $\phi$ increases over time. With this, we calculate the 3+1 decomposition of the kinetic EM tensor to obtain the density and pressure
\begin{equation}\label{rho and P: quintessence}
\begin{aligned}
    \rho_k=u^au^bT_{(k)ab}&=-\frac{1}{2}g^{ab}\partial_a\phi\partial_b\phi = -\frac{1}{2}X\\
    P_k=\frac{1}{3}h^{ab}T_{(k)ab}&=-\frac{1}{2}g^{ab}\partial_a\phi\partial_b\phi = -\frac{1}{2}X,
\end{aligned}
\end{equation}
where $h^{ab}=g^{ab}+u^au^b$ is the projection operator which projects any given tensor living in the 4D spacetime to the 3-hypersurface perpendicular to the DE 4-velocity. It is straightforward to verify that under the definition~\eqref{4-vel 1}, there is no imperfect flow and anisotropic stress in the kinetic EM tensor. Consequently, using quantities defined above, the kinetic EM tensor can be written as
\begin{equation}\label{EM: quintessence fluid}
    T_{(k)ab}=(\rho_k+P_k)u_au_b+P_kg_{ab}.
\end{equation}

One interesting feature of Eq.~\eqref{rho and P: quintessence} is that the formulas of $\rho_k$ and $P_k$ are exactly identical. Since the derivation above neither assumes the scalar field $\phi(x^\mu)$ is homogeneous nor relies on perturbation expansions, this implies that the equation of state for the kinetic part is $P_k=\rho_k$ on all scales. \footnote{It is also straightforward to verify that, for the potential part, we have $\rho_V=-P_V$.}

Now we can derive the fluid equations for quintessence. Using the conservation equation $\nabla_aT^a_{\;\,b}=0$, the second formula in Eq.~\eqref{EM: quintessence}, and Eq.~\eqref{EM relation}, we have
\begin{equation}
\begin{aligned}
    \nabla_aT^{\quad a}_{(k)\;b}&=-\nabla_aT^{\quad a}_{(V)\;b}=\frac{\text{d}V}{\text{d}\phi}\partial_b\phi\\
    &=-\frac{\text{d}V}{\text{d}\phi}\sqrt{-X}u_b=-\sqrt{\rho_k+P_k}\frac{\text{d}V}{\text{d}\phi}u_b.
\end{aligned}
\end{equation}
By decomposing this equation into parallel and perpendicular components with respect to $u^a$ and substituting Eq.~\eqref{EM: quintessence fluid}, we can obtain the fluid equations: $u^b\nabla_aT^{\quad a}_{(k)\;b}=-u^b\nabla_aT^{\quad a}_{(V)\;b}$ leads to the continuity equation
\begin{equation}\label{continuity: quintessence}
    u^a\nabla_a\rho_k+\left(\rho_k+P_k\right)\nabla_au^a=-\sqrt{\rho_k+P_k}\frac{\text{d}V}{\text{d}\phi},
\end{equation}
while $h^{ab}\nabla_cT^{\quad c}_{(k)\;a}=-h^{ab}\nabla_cT^{\quad c}_{(V)\;a}$ gives the Euler equation
\begin{equation}\label{Euler: quintessence}
    \left(\rho_k+P_k\right)u^a\nabla_au^b=-h^{ab}\partial_aP_k.
\end{equation}
In addition, from the definition of 4-velocity, we have
\begin{equation}
    u^a\partial_a\phi=-\frac{\partial^a\phi}{\sqrt{-X}}\partial_a\phi=\sqrt{-X}=\sqrt{\rho_k+P_k}.
\end{equation}
Note that $u^a=\dfrac{\text{d}x^a}{\text{d}\tau}$ with $\tau$ being the proper time along with the motion of DE fluid. As a result, the fluid equations for quintessence are:
\begin{align}
    \frac{\text{d}\textcolor{Coral2}{\rho_k}}{\text{d}\tau}+\left(\textcolor{Coral2}{\rho_k}+\textcolor{Coral2}{P_k}\right)\nabla_a\textcolor{RoyalBlue1}{u^a}&=-\sqrt{\textcolor{Coral2}{\rho_k}+\textcolor{Coral2}{P_k}}\frac{\text{d}V}{\text{d}\textcolor{Green4}{\phi}},\label{continuity2: quintessence}\\
    \left(\textcolor{Coral2}{\rho_k}+\textcolor{Coral2}{P_k}\right)\frac{\text{D}\textcolor{RoyalBlue1}{u^b}}{\text{d}\tau}&=-\left(g^{ab}+\textcolor{RoyalBlue1}{u^a}\textcolor{RoyalBlue1}{u^b}\right)\partial_a\textcolor{Coral2}{P_k},\label{Euler2: quintessence}\\
    \frac{\text{d}\textcolor{Green4}{\phi}}{\text{d}\tau}&=\sqrt{\textcolor{Coral2}{\rho_k}+\textcolor{Coral2}{P_k}}.\label{phi2: quintessence}
\end{align}
Here, for the sake of clear exhibition, we have used different colours to represent different fluid quantities. Along with the EoS $\textcolor{Coral2}{P_k} = \textcolor{Coral2}{\rho_k}$, we now have \textit{six} fluid equations (1 continuity equation, 3 components of the Euler equation, 1 EoS, and 1 for scalar field evolution) with \textit{six} unknown functions (density, pressure, 3 independent components of the 4-velocity, and the scalar field), enabling us to solve this system.

The new fluid equations, Eqs.~\eqref{continuity2: quintessence}-\eqref{phi2: quintessence} are fundamentally distinct from the normal fluid formalism, Eq.~\eqref{normal fluid equations}. First, instead of the total density of the DE fluid, the new equations evolve the kinetic density $\rho_k$ and the scalar field $\phi$. This simplifies the EoS to $w_k=P_k/\rho_k=1$, saving us from the cumbersome perturbation expansion\footnote{This also makes intuitive sense: for standard model particles, we usually treat free particle masses and energy due to interactions separately, and this is the same approach applied to quintessence here.}. The total DE density at any point of spacetime, which is of more interest, can easily be reconstructed as $\rho_{\rm DE}=\rho_k+V(\phi)$. Second, on the right-hand side of the continuity equation, \eqref{continuity2: quintessence}, there is a source term originating from the potential (interacting) part of the theory. This reflects the fact that the kinetic energy density is not only influenced by the change of the effective volume of the fluid element, but will also be infused from or extracted to the potential energy $V(\phi)$, which behaves like a heat reservoir in this case, and the direction of the energy transmission is determined by the sign of $\dfrac{\text{d}V}{\text{d}\phi}$.

Despite these advantages, the sound speed, $c_s$, of quintessence models is always $1$, or the speed of light. Consequently, a small density perturbation of a DE fluid will lead to a very large pressure perturbation, or equivalently, the pressure gradient, preventing the DE fluid from clustering. This renders a nonlinear treatment not strictly necessary for the evolution of quintessence. However, as we will exhibit in the following subsections, the framework established in this subsection can be naturally extended for more complicated DE models, which admit large density contrasts in their evolution and are thus interesting in the research of clustering dark energy.

\subsection{\boldmath \texorpdfstring{$k$}{}-essence}

The action of $k$-essence can be specified as
\begin{equation}
    S=\int \text{d}^4x\sqrt{-g}K(\phi, X).
\end{equation}
By varying the action with respect to $g^{ab}$, we obtain the energy-momentum tensor:
\begin{equation}
    T_{ab}=-2K_X\partial_a\phi\partial_b\phi + g_{ab}K.
\end{equation}
Hereafter, subscript $_\phi$ and $_X$ denotes derivatives w.r.t. $\phi$ and $X$. Through the definition of the 4-velocity, Eq.~\eqref{4-vel 1}, the \textit{total} density and the pressure of DE can be written as
\begin{equation}
\begin{aligned}
    \rho&=2XK_X-K,\\
    P&=K.
\end{aligned}
\end{equation}
Again, there is no imperfect flow or anisotropic stress in this theory. As in quintessence, we define the kinetic part of the $k$-essence to satisfy $\rho_k=P_k$, while for the potential part we have $\rho_V=-P_V$. Through the relation
\begin{equation}
\begin{aligned}
    \rho&=\rho_k+\rho_V,\\
    P&=P_k+P_V,
\end{aligned}
\end{equation}
we have
\begin{equation}\label{rho: k-essence}
\begin{aligned}
    \rho_k&=XK_X,\\
    \rho_V&=XK_X - K,
\end{aligned}
\end{equation}
and the corresponding energy-momentum tensors are
\begin{equation}
\begin{aligned}
    T^{(k)}_{ab} &= (\rho_k+P_k)u_au_b + P_kg_{ab}\\
    T^{(V)}_{ab} &= \left(K-XK_X\right)g_{ab}.
\end{aligned}
\end{equation}

Next, as for quintessence, we can rewrite the conservation equation, $\nabla_bT^{ab}=0$, as
\begin{equation}\label{conservation: k-essence derivation}
\begin{aligned}
    \nabla_bT_{(k)}^{ab}&=-\nabla_bT_{(V)}^{ab}\\
    &=-\left(K_\phi-XK_{\phi X}\right)\nabla^a\phi+XK_{XX}\nabla^aX\\
    &=\left[\sqrt{-X}K_\phi+\left(-X\right)^{3/2}K_{\phi X}\right]u^a\\&\quad\;+XK_{XX}\nabla^aX,
\end{aligned}
\end{equation}
where the last equation is obtained by using Eq.~\eqref{4-vel 1}. We further relate the $\nabla^aX$ to the kinetic density through the covariant derivative of the first relation in Eq.~\eqref{rho: k-essence}:
\begin{equation}
\begin{aligned}
    \nabla^a\rho_k&=\left(K_X+XK_{XX}\right)\nabla^aX+XK_{\phi X}\nabla^a\phi\\
    &=\left(K_X+XK_{XX}\right)\nabla^aX+\left(-X\right)^{3/2}K_{\phi X}u^a,
\end{aligned}
\end{equation}
which leads to
\begin{equation}
    \nabla^aX=\frac{\nabla^a\rho_k-\left(-X\right)^{3/2}K_{\phi X}u^a}{K_X+XK_{XX}}.
\end{equation}
Substituting this into the last line of Eq.~\eqref{conservation: k-essence derivation} yields
\begin{equation}
\begin{aligned}
    \nabla_bT_{(k)}^{ab}&=\left[\sqrt{-X}K_\phi+D(\phi, X)\left(-X\right)^{3/2}K_{\phi X}\right]u^a\\&\quad\;+\left[C(\phi, X) - 1\right]\nabla^a\rho_k,
\end{aligned}
\end{equation}
where we have defined two factors
\begin{equation}
\begin{aligned}
    C(\phi, X)&=\frac{K_X(\phi, X)+2XK_{XX}(\phi, X)}{K_X(\phi, X)+XK_{XX}(\phi, X)},\\
    D(\phi, X)&=\frac{K_X(\phi, X)}{K_X(\phi, X)+XK_{XX}(\phi, X)}.
\end{aligned}
\end{equation}
Using the definition of the sound speed squared,
\begin{equation}
    c_s^2(\phi, X)=\frac{K_X(\phi, X)}{K_X(\phi, X)+2XK_{XX}(\phi, X)},
\end{equation}
these factors above can be represented as
\begin{equation}
\begin{aligned}
    C(\phi, X)&=\frac{2}{1+c_s^2(\phi, X)},\\
    D(\phi, X)&=\frac{2c_s^2(\phi, X)}{1+c_s^2(\phi, X)}.
\end{aligned}
\end{equation}
\vspace{-2pt}

We subsequently decompose the conservation equation with respect to the 4-velocity by expanding the formula $u_a\nabla_bT^{ab}=0$ and $h^a_c\nabla_bT^{bc}=0$. This leads to the continuity equation
\begin{equation}
\begin{aligned}
    &\quad C\,u^b\nabla_b\rho_k + \left(\rho_k+P_k\right)\nabla^au_a\\ &= \sqrt{-X}K_\phi + D\left(-X\right)^{3/2}K_{\phi X},
\end{aligned}
\end{equation}
and the Euler equation
\begin{equation}
    \left(\rho_k+P_k\right)u^b\nabla_bu^a = - D\left(\rho_k, \phi\right)h^{ab}\nabla_bP_k.
\end{equation}
At the same time, the evolution of the scalar field is given by
\begin{equation}
    u^a\partial_a\phi=-\frac{\partial^a\phi}{\sqrt{-X}}\partial_a\phi=\sqrt{-X(\rho_k, \phi)}.
\end{equation}
In the above equations, $X=X(\rho_k, \phi)$ can be calculated by inverting 
$\rho_k=XK_X(\phi, X)$. 

Dividing $C(\phi, X)$ on both sides of the continuity and Euler equations, the full fluid equations of $k$-essence can be simplified as
\begin{widetext}
\begin{align}
    \frac{\text{d}\textcolor{Coral2}{\rho_k}}{\text{d}\tau}+\left[\textcolor{Coral2}{\rho_k}+\textcolor{Coral2}{P_k}c_s^2\left(\textcolor{Green4}{\phi}, X\right)\right] \nabla_a\textcolor{RoyalBlue1}{u^a}&=\frac{1+c_s^2\left(\textcolor{Green4}{\phi}, X\right)}{2}\sqrt{-X\left(\textcolor{Coral2}{\rho_k}, \textcolor{Green4}{\phi}\right)}K_{\textcolor{Green4}{\phi}} + c_s^2\left(\textcolor{Green4}{\phi}, X\right)\left(-X\left(\textcolor{Coral2}{\rho_k}, \textcolor{Green4}{\phi}\right)\right)^{3/2}K_{\textcolor{Green4}{\phi}X},\label{continuity2: k-essence}\\
    \left[\textcolor{Coral2}{\rho_k}+\textcolor{Coral2}{P_k}c_s^2\left(\textcolor{Green4}{\phi}, X\right)\right]\frac{\text{D}\textcolor{RoyalBlue1}{u^a}}{\text{d}\tau}&=-c_s^2\left(\textcolor{Green4}{\phi}, X\right)\left(g^{ab}+\textcolor{RoyalBlue1}{u^a}\textcolor{RoyalBlue1}{u^b}\right)\partial_b\textcolor{Coral2}{P_k},\label{Euler2: k-essence}\\
    \frac{\text{d}\textcolor{Green4}{\phi}}{\text{d}\tau}&=\sqrt{-X\left(\textcolor{Coral2}{\rho_k}, \textcolor{Green4}{\phi}\right)}.\label{phi2: k-essence}
\end{align}
\end{widetext}
By solving the equation $\textcolor{Coral2}{\rho_k}=XK_X\left(\textcolor{Green4}{\phi}, X\right)$ to obtain $X=X\left(\textcolor{Coral2}{\rho_k}, \textcolor{Green4}{\phi}\right)$, and combining the EoS $\textcolor{Coral2}{\rho_k}=\textcolor{Coral2}{P_k}$, we can close the above system. Note that in the above equations we have used the EoS to simplify
\begin{equation}
\begin{aligned}
    &\frac{1+c_s^2}{2}\left(\rho_k+P_k\right)=\left(1+c_s^2\right)\rho_k
    \\=&\rho_k+\rho_k c_s^2=\rho_k+P_k c_s^2.
\end{aligned}
\end{equation}
Though it essentially doesn't change the fluid equations, this form provides us with a clearer look at how the influence of the DE (kinetic) pressure is affected by the sound speed squared.

Before moving on, it is worth commenting on the factor $c_s^2(\phi, X)$ appearing in the above equations. If we substitute the background value of the scalar field, namely $\bar{\phi}(t)$ and $\bar{X}(t)$, $c_s^2(\bar{\phi}, \bar{X})$ becomes the background sound speed squared $\bar{c}^2_s(t)$ for the DE model, which is commonly employed in perturbation theory. In general, however, the scalar field can deviate from the background value, making the sound speed squared of the DE fluid deviate from its background value as well (even a small perturbation in $\phi$ may lead to strong deviation of $c_s^2$ from its background value due to the strongly nonlinear relation between the two). 

Another interesting feature of $c_s^2$ is that the (kinetic) pressure always appears with the $c_s^2$ in Eqs.~\eqref{continuity2: k-essence} and~\eqref{Euler2: k-essence}. In the limit $c_s^2\rightarrow0$, both the $c_s^2P_k$ and the gradient term $c_s^2h^{ab}\partial_bP_k$ tend to zero, weakening the resistance of the pressure gradient on DE clustering. Consequently---and as expected---with a lower sound speed, the dark energy fluid will behave increasingly like pressureless dark matter, though their background evolutions are still different.

\section{Application to Spherical Collapse}\label{sec: SC}

As preliminary examples, we apply the fluid equations in the last section to a set of spherical collapse simulations. We consider two different $k$-essence models. The first is called the \textit{XE model} (which stands for ``X Expansion"), whose Lagrangian density is written as
\begin{equation}
    K(\phi, X)=V(\phi)+f(\phi)X+q(\phi)X^2.
\end{equation}
The second is the \textit{SV model} (which stands for ``Separating Variables"), whose Lagrangian function is
\begin{equation}
    K(\phi, X)=q(\phi)h(X).
\end{equation}

In the simulations, we use the CPL parametrization \cite{CPL01, CPL02} 
of the equation of state of the DE fluid, 
\begin{equation}
    w(a)=w_0+w_a(1-a),
\end{equation}
where $a$ is the scale factor. For concreteness, we adopt $w_0=-0.8$ and $w_a=-0.1$ for both 
models. We fix the background values of their sound speeds squared,
\begin{equation}
    \bar{c}_s^2(t)=\frac{K_X(\bar{\phi}, \bar{X})}{K_X(\bar{\phi}, \bar{X})+2XK_{XX}(\bar{\phi}, \bar{X})},
\end{equation}
to $10^{-6}$ for the fiducial simulation, but we will show results of other values to see the effect of varying $\bar{c}^2_s$. 
In addition, we assume the present-day density parameters of DM and DE to be $\Omega_m=0.3$ and $\Omega_{\text{DE}}=0.7$. As a result, the XE and SV models will have identical background and linear evolutions. 

In the following subsections, we will briefly introduce how to reconstruct the functional form of these two models using their EoS and background $c_s^2$, present the equations used to evolve the density, velocity, scalar field, and gravity in spherical collapse simulations, and analyze 
the non-linear evolution of these two models at different redshifts.

\begin{figure*}[t]
\includegraphics[width=\textwidth]{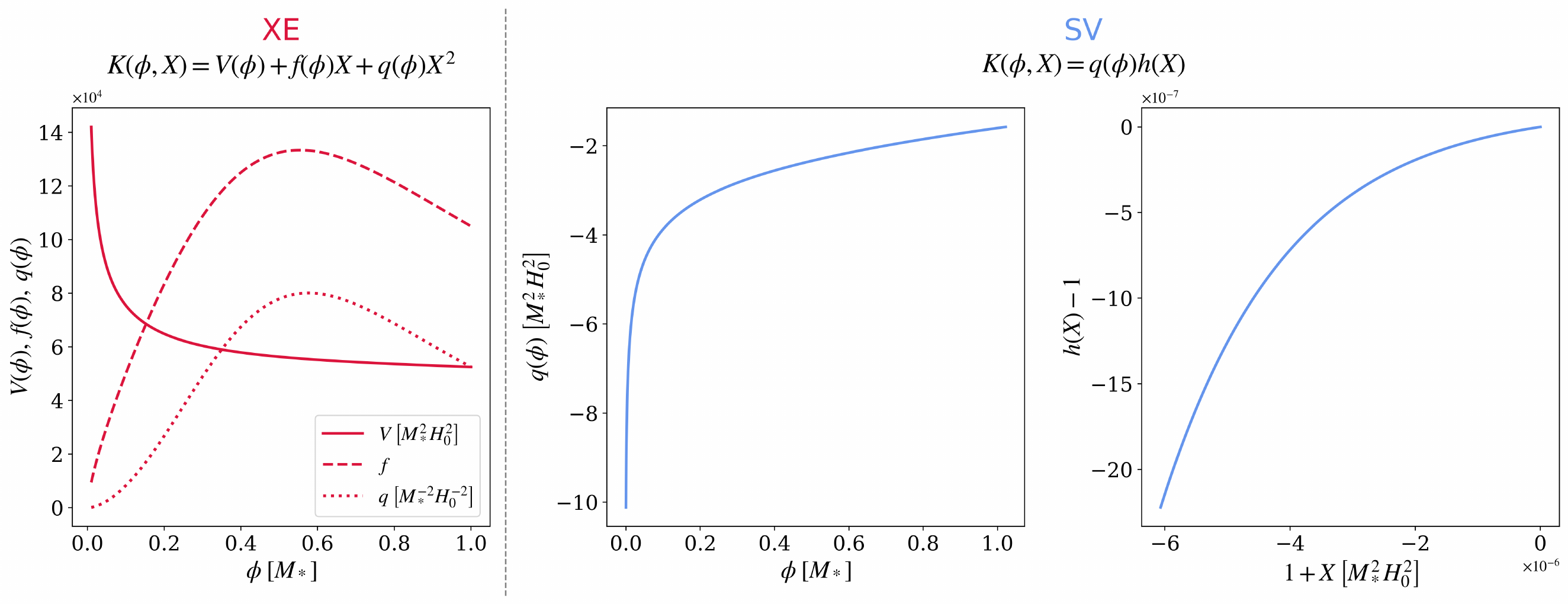}
\caption{Reconstruction results of DE models with $w_0 = -0.8,\; w_a = -0.1,\; \bar{c}_s^2=10^{-6}$, $\Omega_m=0.3$ and $\Omega_{\text{DE}}=0.7$. \textit{First panel:} Reconstructed $V(\phi)$ (solid line, normalized by $M_*^2H_0^2$), $f(\phi)$ (dashed line), and $q(\phi)$ (dotted line, normalized by $M_*^{-2}H_0^{-2}$) against $\phi$ (normalized by $M_*$) in the XE model. \textit{Second and Third panels:} Reconstructed $q(\phi)$ versus $\phi$ and $h(X)$ versus $X$ in the SV model, with $q$, $\phi$, and $X$ normalized by $M_*^2H_0^2$, $M_*$, and $M_*^2H_0^2$ respectively.}
\label{fig:reconstruction}
\end{figure*}

\subsection{Brief Review of the DE Reconstruction}\label{subsec: reconstruction}
This section is a succinct summary of the reconstruction method in \cite{Reconstruction}, and is mainly focused on the $k$-essence models. We refer readers to this paper for more general and detailed derivations. 

The starting point of the reconstruction is the effective field theory of dark energy (EFTofDE) (see \cite{EFTofDE} as a review), whose action (for $k$-essence models) can be written as
\begin{equation}
\begin{aligned}
    S_{\text{EFT}}=\int\text{d}^4x\sqrt{-g}&\Bigg[\frac{M_*^2}{2}R-\Lambda(t)-c(t)g^{00}\\
    &+\frac{M_2^4(t)}{2}\left(g^{00}\right)^2+\cdots\Bigg],
\end{aligned}
\end{equation}
where $M_*$ is the reduced Planck mass, $g^{00}$ is the time-time component of the metric $g^{ab}$, and the time-dependent functions, $\Lambda(t)$, $c(t)$, and $M_2^4(t)$, also known as the EFT parameters,  are related to the background quantities of the $k$-essence model by
\begin{equation}\label{EFT relations}
\begin{aligned}
    c(t)-\Lambda(t)&=K(t),\\
    c(t)&=X(t)K_X(t),\\
    M_2^4(t)&=X(t)^2K_{XX}(t).
\end{aligned}
\end{equation}
We can also connect the three EFT parameters with some physical quantities of the DE:
\begin{equation}\label{EFT relations with fluid quantities}
\begin{aligned}
    c(t)+\Lambda(t)&=\rho_{\text{DE}}(t)\\
    c(t)-\Lambda(t)&=P_{\text{DE}}(t)\\
    M_2^4(t)&=\frac{1}{2}\left(\frac{1}{\bar{c}_s^2(t)}-1\right)c(t).
\end{aligned}
\end{equation}
Therefore, as long as we have specified the related physical parameters (in our case they are $\Omega_m=0.3$, $\Omega_{\text{DE}}=0.7$, $w_0=-0.8$, $w_a=-0.1$, and $\bar{c}_s^2(t)\equiv10^{-6}$ by default), we can obtain the time evolution of the three EFT parameters. Substituting them into Eq.~\eqref{EFT relations}, it becomes a set of equations for $K=K(t)$, $\phi=\phi(t)$, and $X=X(t)$. 

For the XE model, if we take advantage of the field redefinition to make $\phi(t) = a(t)$, Eq.~\eqref{EFT relations} simply becomes
\begin{eqnarray}
    c(a)-\Lambda(a) &=& V(a)-f(a)\left(aH\right)^2+q(a)\left(aH\right)^4,\nonumber\\
    c(a) &=& -f(a)\left(aH\right)^2 + 2q(a)\left(aH\right)^4,\\
    M_2^4(a) &=& 2q(a)\left(aH\right)^4.\nonumber
\end{eqnarray}
We can solve these equations and obtain $V(a)$, $f(a)$, and $q(a)$, and inverse $\phi=\phi(a)$ to give $a = a(\phi)$. Substituting it into $V(a)$, $f(a)$, and $q(a)$ will lead to $V(\phi)$, $f(\phi)$, and $q(\phi)$. Likewise, for the SV model, Eq.~\eqref{EFT relations} can be transformed into
\begin{equation}
\begin{aligned}
    q&=\frac{c-\Lambda}{h},\\
    \frac{\dot{h}}{h}&=\frac{c}{c-\Lambda}\frac{\dot{X}}{X},\\
    \frac{\dot{X}}{X}&=\left(c-\Lambda\right)\frac{\textrm{d}}{\textrm{d}t}\left[\frac{c}{c-\Lambda}\right]\bigg/\left(M_2^4-\frac{c\Lambda}{c-\Lambda}\right).
\end{aligned}
\end{equation}
Solving these equations gives us $q(t)$, $h(t)$ and $X(t)$, while $\phi(t)$ can be obtained from the integration of $\sqrt{-X(t)}$. Subsequently, inverting $\phi(t)$ and $X(t)$ and substituting them into $q(t)$ and $h(t)$ will lead to $q(\phi)$ and $h(X)$. 

The reconstructed XE and SV models used in our fiducial simulations are displayed in Figure.~\ref{fig:reconstruction}. As we chose a very small background sound speed squared in the reconstruction process, the DE models exhibited in this figure are highly nonlinear. For the XE model, 
$V$, $f$, and $q$ are nonlinear functions of order $\mathcal{O}(10^4-10^5)$. When we add them together to calculate $K$ or $XK_X$, these three functions will mostly cancel out 
and lead to $\mathcal{O}(0.1-1)$ results, but 
the corresponding $K_{XX}$ 
is $\mathcal{O}(10^5)$, 
much greater than $K$ and its first-order derivative with respect to $X$. 
The SV model is rather different: the $q$ function is of the order of $\mathcal{O}(1)$ and the nonlinearity mainly lies in $h(X)$. It is not easy to notice this feature in Fig.~\ref{fig:reconstruction}, but it can be shown that, partly because $X$ only varies in a very narrow range in this case, the derivative $|h_X|$ is of order $\mathcal{O}(1)$, which is far larger than $|h-1|$, and $|h_{XX}|$ is of order $\mathcal{O}(10^6)$, which is in turn far larger than $|h_X|$. This strong enhancement of $K_{XX}$ relative to $K_X$ represents one of the key features distinguishing the clustering dark-energy models considered here from the non-clustering ones.

\subsection{Spherical-Collapse Equations}
\label{SC equations}

We formulate the spherical-collapse calculation in physical coordinates, which provide a natural description of the evolution of the physical radii of the matter and dark-energy shells. This choice is also consistent with the conventional formulation of spherical collapse for collisionless dark matter
. During the validation and debugging of our numerical scheme, we also reran our code in comoving coordinates and compared the two. 
The results fully agree with each other. For the sake of clarity, we write the speed of light $c$ explicitly in the following part of this subsection.  

The simulations were done under the 
Newtonian gauge, for which the metric can be written as
\begin{equation}
    \text{d}s^2=-\left(1+2\Psi/c^2\right)\text{d}t^2+a^2\left(1-2\Phi/c^2\right)\delta_{ij}\text{d}x^i\text{d}x^j.
\end{equation}
The continuity equation can be derived from Eq.~\eqref{continuity2: k-essence} as
\begin{equation}\label{SC continuity}
\begin{aligned}
    &\frac{\text{d}\rho_k}{\text{d}t}+\rho_k\left(1+\frac{c_s^2}{c^2}\right)\frac{1}{r^2}\frac{\partial}{\partial r}\left(r^2u_{\text{DE}}\right)\\
    =&\frac{1}{2}\left(1+\frac{c_s^2}{c^2}\right)\sqrt{-X}K_{\phi}+\frac{c_s^2}{c^2}(-X)^{3/2}K_{X\phi}, 
\end{aligned}
\end{equation}
where 
\begin{equation}
    u_{\text{DE}} = H(z)r+v_{r}
\end{equation}
is the physical total radial velocity (the sum of the Hubble flow at redshift $z$ and physical radius $r$ and the physical peculiar velocity).

In the spherical-collapse simulations, we discretize the whole space into a series of concentric thin shells. The $i$-th shell has mass $m_{k,i}=\rho_{k,i}V_i$ and volume $V_i$. Through the relationship
\begin{equation}
    \frac{\text{d}V}{\text{d}t}=V\nabla\cdot\Vec{u}=V\nabla_ru_{\text{DE}},
\end{equation}
we can derive the time derivative of the mass inside the $i$-th shell:
\begin{equation}
\begin{aligned}
    &\;\frac{\text{d}m_{k,i}}{\text{d}t}=V_i\frac{\text{d}\rho_{k,i}}{\text{d}t}+\rho_{k,i}\frac{\text{d}V_i}{\text{d}t}\\
    =&\;\frac{1}{2}V_i\left[\left(1+\frac{c_{s,i}^2}{c^2}\right)\sqrt{-X_i}K_{\phi,i}+\frac{c_{s,i}^2}{c^2}(-X_i)^{3/2}K_{X\phi,i}\right]\\
    &-\frac{c_{s,i}^2}{c^2}m_{k,i}\frac{\textrm{d}\ln{V_i}}{\textrm{d}t}.
\end{aligned}
\end{equation}

Similarly, by subtracting the time component of Eq.~\eqref{Euler2: k-essence} times the peculiar velocity from the spatial component of the same equation, we obtain the Euler equation for spherical collapse:
\begin{equation}\label{SC Euler}
    \frac{\textrm{d}u_{\textrm{DE}}}{\textrm{d}t}=-\nabla_r\Psi_N-\dfrac{c_s^2\nabla_r\rho_k}{\rho_k(1+c_s^2/c^2)},
\end{equation}
with the Newtonian gravitational potential, $\Psi_N$ \footnote{Note that as these equations are in the physical coordinates, this $\Psi_N$ is different from the gravity perturbation $\Psi$ in Appendix~\ref{app: fluid and poisson}. More specifically, $\Psi_N$ is computed by $\rho+3P$, while $\Psi$ in Appendix~\ref{app: fluid and poisson} is computed by $\delta \rho+3\delta P$}, computed using the Poisson equation (under spherical symmetry):
\begin{equation}\label{SC Poisson}
\begin{aligned}
    \frac{1}{r^2}\frac{\text{d}}{\text{d}r}\left(r^2\frac{\text{d}\Psi_N}{\text{d}r}\right)&=4\pi G\left(\rho_{\text{DM}}+\rho_{\text{DE}}+3\frac{P_{\rm DE}}{c^2}\right)\\
    &=4\pi G\left(\rho_{\text{DM}}+2\rho_{k}+2K(\phi, X)\right).
\end{aligned}
\end{equation}

Finally, the evolution equation for $\phi$ is
\begin{equation}\label{SC phi}
    \frac{\text{d}\phi}{\text{d}t}=\sqrt{1+2\frac{\Psi}{c^2}-\frac{v_r^2}{c^2}}\sqrt{-X},
\end{equation}
with $X=X\left(\rho_k, \phi\right)$ being solved from $\rho_k=XK_X(\phi, X)$. For the XE model, $X$ can be derived from
\begin{equation}
    X=\frac{-f(\phi)-\sqrt{f(\phi)^2+8\rho_kq(\phi)}}{4q(\phi)},
\end{equation}
while for the SV model, $X$ is numerically solved using
\begin{equation}
\begin{aligned}
    X&=J^{-1}\left[\rho_k/q(\phi)\right],\\
    J(X)&=Xh_X(X).
\end{aligned}
\end{equation}

It is worth noting that, in the fluid equations~\eqref{SC continuity}~\eqref{SC Euler} and the Poisson equation~\eqref{SC Poisson}, we have neglected most of the terms that are related to the gravitational potential, such as $\Psi$ and $\dot{\Phi}$. These terms correspond to relativistic corrections on large linear scales, while for our spherical collapse simulation, which mainly focuses on the scale of a few Mpc, they have little influence on the final results. However, in the evolution equation of $\phi$, Eq.~\eqref{SC phi}, we have kept the $2\Psi$ term, as well as the second-order term $v_{r}^2 / c^2$, because in this case they do play a significant role in the ``velocity reconstruction" from the gradient of $\phi$. A detailed derivation and a more comprehensive discussion of these tricky points are given in Appendix~\ref{app: fluid and poisson}.

For the sake of completeness, we also list the continuity and Euler equations for the DM component:
\begin{equation}
\begin{aligned}
    \frac{\textrm{d}\rho_{\rm DM}}{\textrm{d}t}+\frac{1}{r^2}\frac{\partial}{\partial r}\left(r^2u_{\text{DM}}\right)&=0\\
    \frac{\textrm{d}u_{\rm DM}}{\textrm{d}t}&=-\nabla_r\Psi_N,
\end{aligned}
\end{equation}
which have been extensively used in DM-only spherical collapse simulations.

\subsection{Numerical Setup and Initial Conditions}

\begin{figure*}[t]
\includegraphics[width=\textwidth]{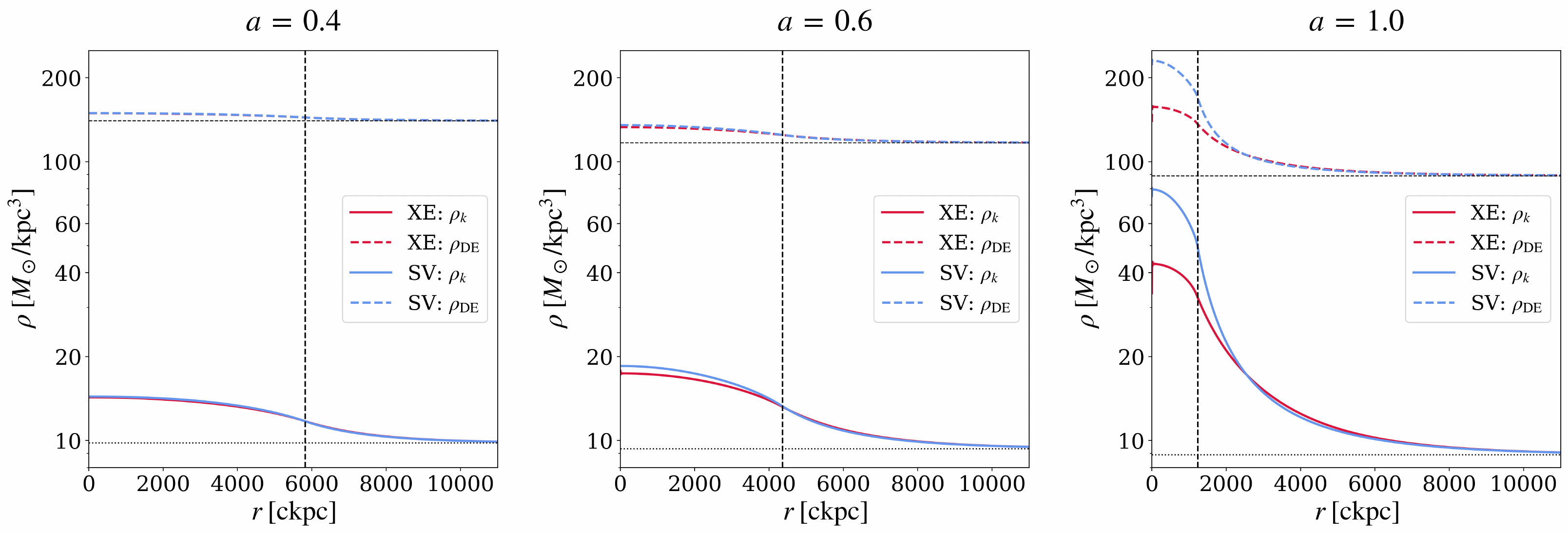}
\caption{The DE density profiles against the \textit{comoving} radius at three times: $a=0.4$ (left panel), $a=0.6$ (middle), and $a=1.0$ (right). In each panel, we present the profiles of the kinetic energy density $\rho_k=XK_X$ (solid lines) and the total energy density $\rho_{\rm DE}=2XK_X-K$ (dashed lines). Red and blue colors indicate, respectively, the XE and SV models. The vertical dashed line marks the average radius of the DM halo (the innermost 1000 shells) in the two models, since the difference between the radius of the DM halo in these models is so small that they will overlap with each other. The background values of the kinetic density (horizontal dotted line) and the total DE density (horizontal dashed line) at each redshift are also plotted for reference.
}
\label{fig:density profiles}
\end{figure*}

We consider a flat universe containing two components, dark matter and dark energy. The cosmological parameters adopted in the simulations are $\Omega_{\rm m}=0.3$ and $\Omega_{\rm DE}=0.7$, with the dark-energy equation-of-state parameters fixed to $w_0=-0.8$ and $w_a=-0.1$. For the fiducial simulations, we adopt the background value of the dark-energy sound speed to be $\bar{c}_s^2\equiv 10^{-6}$. These parameters have been introduced in the preceding subsections, but we list them again here for convenience and to make the numerical setup self-contained. The present-day Hubble parameter is set to $H_0=67.66{\rm \;km\cdot s^{-1}\cdot Mpc^{-1}}$.

The simulation is initialized at $z_{\rm ini}=49$ and followed until $z=0$. Both the dark-matter and dark-energy components are discretized into 3000 concentric spherical shells. The time integration is performed using a uniform spacing in $a^{3/2}$, with $\Delta a^{3/2}=2.5\times10^{-6}$ for each time step. The spatial gradient terms in the fluid equations presented in the last subsection are calculated through the finite difference method. At the initial time, the velocities of both dark-matter and dark-energy shells are set to the local Hubble-flow velocity corresponding to their initial radii. For DM, we adopt a spherical top-hat density profile. The innermost DM 1000 shells are assigned a uniform overdensity of $\delta_{\rm DM}=0.076$ relative to the background, while the remaining 2000 shells are initially unperturbed. More precisely,
\begin{equation}\label{initial tophat}
    \delta_{\rm DM, ini}(r)=\left\{
    \begin{array}{cc}
        0.076 & \quad r\leq R_{\rm halo,ini} \\
        0 & \quad r > R_{\rm halo,ini},
    \end{array}
    \right.
\end{equation}
where $R_{\rm halo,ini}\approx167{\rm \;kpc}$ is the radius of the 1000-th shell. In the fiducial setup, the innermost 1000 shells contain a total dark-matter mass of $10^{14} M_\odot$, which is the mass of the dark-matter halo. The chosen overdensity is such that this region approximately reaches one half of the turnaround radius of the DM halo by $z=0$. For DE, we set the initial density to be the background DE density for all the 3000 shells, or specifically, $\delta_{\rm DE, ini}(r)\equiv 0$. Outside the simulated region, the dark-matter and dark-energy distributions are assumed to evolve consistently with the cosmological background.

In addition to the fiducial simulations, we also explore the dependence of DE clustering on the background sound speed and DM halo mass by varying $\bar{c}_s^2$ ($10^{-4}$, $10^{-6}$, and $10^{-8}$) and $M_{\rm DM}$ ($10^{13}M_\odot$, $10^{14}M_\odot$, and $10^{15}M_\odot$), respectively, while keeping the remaining cosmological parameters and initial conditions fixed.

We stress that the initial condition adopted here is not intended to serve as a physical cosmological initial condition. Rather, the primary purpose of the spherical-collapse calculation is to demonstrate an application of the nonlinear fluid equations derived in the preceding sections. Therefore, we choose such an initial condition for the sake of simplicity. A more realistic choice of initial conditions would modify the detailed evolution of the system, but we expect it not to qualitatively alter the main results presented in the following section, or the discussion in a companion paper \cite{CompanionPRL}. 

\subsection{Simulation Result}

As stated in the last subsection, we stop our spherical collapse simulations at $z=0$, when the radius of the DM halo (the innermost 1000 shells) is slightly smaller than half of its turnaround radius. At this stage, the density profile of the dark matter remains approximately a top-hat profile, and the central density of the DM halo is a few hundred times higher than the background DM density, which is close to the density contrast of the DM halos in the 3D cosmological simulations. To illustrate the process of DE clustering, we display the DE density profiles at $a=0.4$, $0.6$, and $1.0$ in Figure.~\ref{fig:density profiles}. 


As the collapse proceeds, dark energy develops increasingly pronounced density perturbations in the vicinity of the dark-matter halo. At $a=0.4$, the density profiles exhibit only modest deviations from the homogeneous background, whereas the central overdensities become progressively more significant at later times. By $a=1.0$, substantial dark-energy clustering has developed within and around the collapsed halo, while the density profiles gradually approach their background values at sufficiently large radii. These results illustrate the ability of our fluid framework to follow the spatial evolution of clustering dark energy into the nonlinear regime. The numerical convergence of the simulations is examined in Appendix~\ref{app: convergence}.

\begin{figure*}[t]
\includegraphics[width=\textwidth]{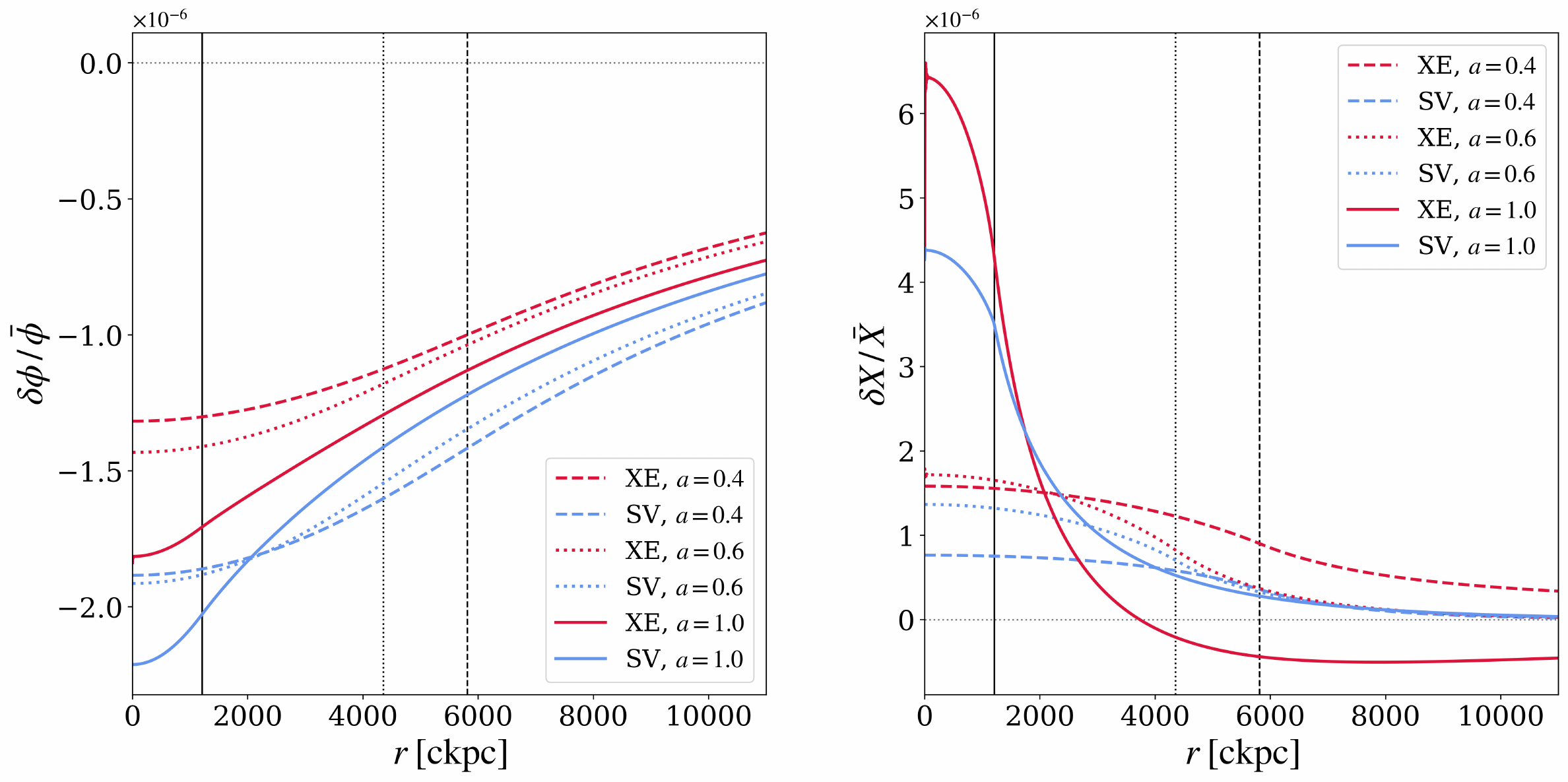}
\caption{Profiles of the scalar perturbations, $\delta\phi/\bar{\phi}$ (left panel) and $\delta X/\bar{X}$(right panel), at $a=0.4$ (dashed lines), $a=0.6$ (dotted lines), and $a=1.0$ (solid lines) for XE (red) and SV (blue) models. The x-axes are the \textit{comoving} radius. 
We 
mark the average comoving DM-halo radius for XE and SV models at $a=0.4$ (vertical dashed lines), $a=0.6$ (vertical dotted lines), and $a=1.0$ (vertical solid lines).}
\label{fig:delta_phi_and_X}
\end{figure*}

\begin{figure*}[t]
\includegraphics[width=\textwidth]{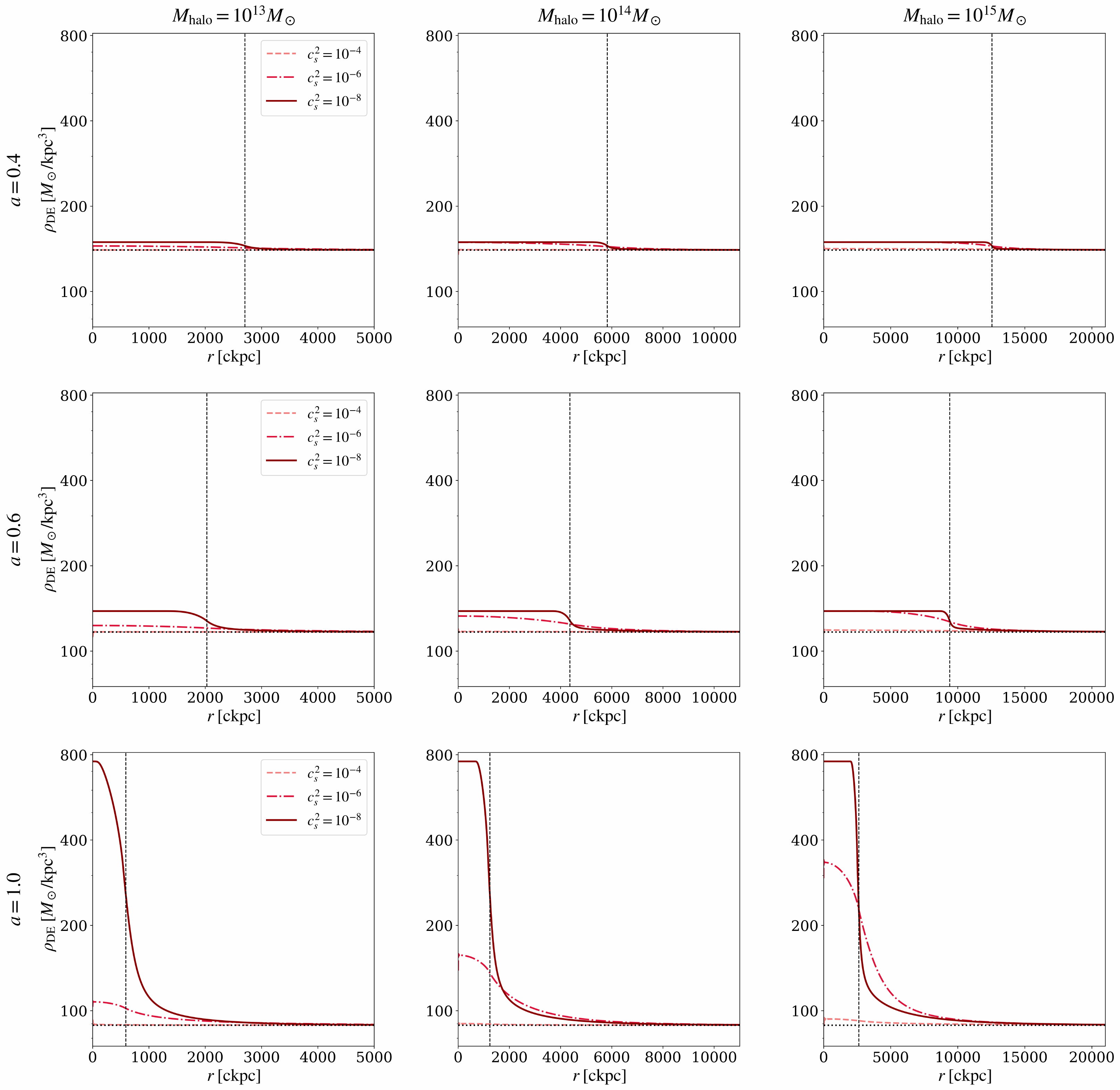}
\caption{The total DE density profiles 
for the XE model at $a=0.4$ (first row), $a=0.6$ (second row), and $a=1.0$ (third row), with the mass of the DM halo being $10^{13}M_\odot$ (first column), $10^{14}M_\odot$ (second column), and $10^{15}M_\odot$ (third column), and with $c_s^2=10^{-4}$ (light red dashed line), $c_s^2=10^{-6}$ (red dashed-dotted line), and $c_s^2=10^{-8}$ (dark red solid line). For each 
halo mass and scale factor, the vertical dashed line marks the radius of the DM halo corresponding to the $c_s^2=10^{-6}$ model. We also plot the background value of the total DE density (horizontal dotted line) at each redshift.
}
\label{fig:Mass_and_cs2_XE}
\end{figure*}

\begin{figure*}[t]
\includegraphics[width=\textwidth]{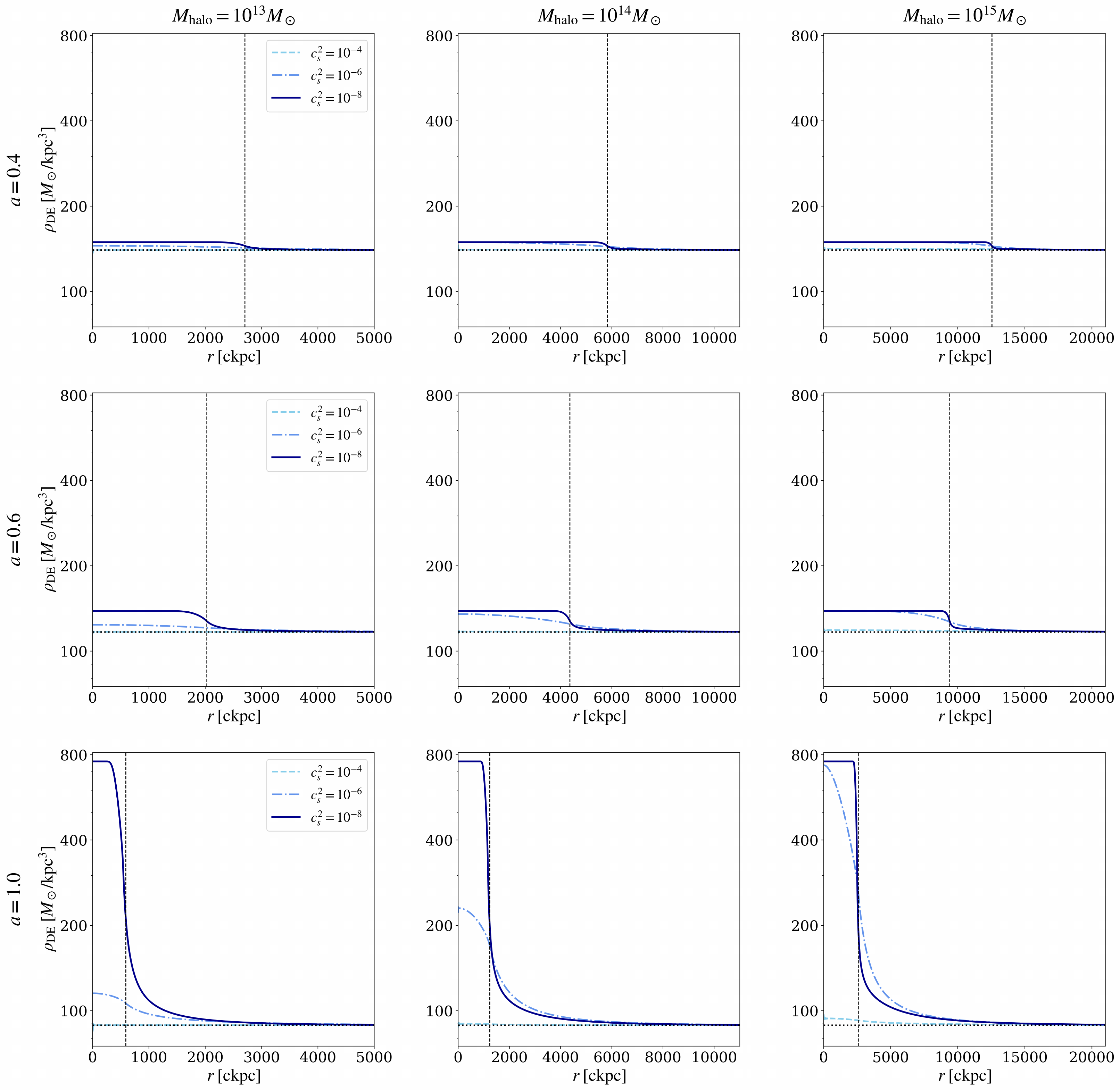}
\caption{Same as Fig.~\ref{fig:Mass_and_cs2_XE}, but for the SV model.
}
\label{fig:Mass_and_cs2_SV}
\end{figure*}

We also present the radial profiles of $\delta\phi/\bar{\phi}$ and $\delta X/\bar{X}$ at $a=0.4$, $0.6$, and $1.0$ for both models in Figure~\ref{fig:delta_phi_and_X}. 
At both redshifts, the perturbations in $\phi$ and $X$ remain at the level of $\sim\mathcal{O}(10^{-6})$, while the corresponding dark-energy density perturbations can reach $\mathcal{O}(1)$, which is a well-known feature of clustering $k$-essence models. It is worth noting that, in the left panel, the perturbations in $\phi$ do not converge to zero (the background value) at large radii. This is caused by the use of Eq.~\eqref{SC phi} as the evolution equation for $\phi$, in which the negative gravitational potential will decrease the value of the scalar field compared to its background value. As we 
have a tophat profile for DM and a constant background profile for DE as initial conditions, the average density in our simulation region is higher than the background density, making the gravitational potential, $\Psi$, 
negative over all of the radius. This thereby reduces the value of $\phi$ for shells at all radii in our simulation compared to the background value. 

One may also notice that, in the right panel of Fig.~\ref{fig:delta_phi_and_X}, $\delta X/\bar{X}$ for the XE model does not return to zero at large radii. This is as expected too, since we solve $X$ from $\rho_k=XK_X(\phi, X)$ in simulations. At large radii, the density converges to its background value, but $\phi$ does not. For the equation to hold, $X$ must also deviate from its background value. For the XE model, $f$ and $q$ are $\mathcal{O}(10^{4})$, considerably amplifying the influence of the deviation of $\phi$, while for the SV model, its $q$ is of order $\mathcal{O}(1)$, making $q(\bar{\phi}+\delta \phi)\approx\bar{q}$, and making $Xh_X(X)\approx\rho_k/\bar{q}$ and thus $X$ converge to background values. In 
three-dimensional simulations, however, the average total density is usually set to 
the background density, so we can expect that this feature will not appear. 

To further explore the applicability of our numerical framework under different physical conditions, we investigate the dependence of DE clustering on the background sound speed squared and the DM halo mass. Specifically, we consider three values of $\bar{c}_s^2=10^{-4},10^{-6},10^{-8}$ (the corresponding models are all reconstructed using the method in Subsection~\ref{subsec: reconstruction}), and three halo masses, $M_{\rm halo}=10^{13},10^{14},10^{15}M_\odot$. The resulting DE density profiles at $a=0.4$, $0.6$, and $1.0$ are presented in Figs.~\ref{fig:Mass_and_cs2_XE} and~\ref{fig:Mass_and_cs2_SV}. 

A clear dependence on the background sound speed can be found in both models. For a given halo mass, decreasing $\bar{c}_s^2$ 
leads to stronger DE clustering. For $\bar{c}_s^2=10^{-4}$, the DE density remains close to its 
background value, with only tiny perturbations developing throughout the evolution. In contrast, for $\bar{c}_s^2=10^{-8}$, the density profiles exhibit a nearly flat central region followed by a 
sharp decline near the DM halo boundary, particularly at late times. In addition, for both models (and all halo masses), these low sound speed DE models achieve almost identical central densities, namely $\sim 760M_\odot/{\rm kpc}^3$. These are reasonable, as discussed at the end of Section~\ref{sec: equation derivation}, since on the right-hand side of the Euler equation, Eq.~\eqref{Euler2: k-essence}, the gradient of the kinetic pressure is multiplied by a pre-factor $c_s^2$, which controls the effectiveness of the pressure-gradient force that prevents gravitational collapse. With a low sound speed, the pressure support becomes negligible, allowing the DE to collapse nearly unimpeded and its density distribution to more closely follow the dark matter. Also, the initial conditions and the collapse time for all the simulations are identical, leading to the same central DE densities.

The halo mass also plays an important role in determining the spatial extent and amplitude of DE clustering. For a fixed background sound speed, more massive halos 
produce stronger density perturbations, 
over larger radial distances, of the DE fluid. This dependence is particularly evident for $\bar{c}_s^2=10^{-6}$, where the central overdensity becomes progressively more pronounced as the halo mass increases. For $\bar{c}_s^2=10^{-8}$, despite nearly identical central overdensity, the central plateau becomes wider and the density transition at the boundary of the plateau becomes sharper for a larger halo mass. These results illustrate how the interplay between pressure support and gravitational force impacts the spatial distribution of clustering DE.

Taken together, these simulations, with the convergence tests in Appendix \ref{app: convergence}, demonstrate that our fluid framework can follow the nonlinear evolution of clustering DE for different sound speeds and gravitational environments.

We note that simulations with larger background sound speeds, such as $\bar{c}_s^2=10^{-3}$, develop numerical oscillations under the present discretization and time-stepping settings, preventing stable evolution to $z=0$. This limitation is associated with the numerical treatment of stronger pressure-gradient forces and does not necessarily indicate a physical instability of the DE fluid. This is, however, not a problematic issue for cosmological simulations. Since $\bar{c}_s^2=10^{-4}$ will only lead to a small perturbation in the DE component, we can expect that the influence of a DE fluid with a larger $\bar{c}_s^2$ is smaller and more linear. Therefore, in these regimes, linear solvers should already accurately predict the evolution of the DE fluid.

\section{Discussions and Conclusions}\label{sec: discussion and conclusion}

We presented a new fluid method for simulating clustering dark energy through 
action-splitting, 
where we divide the action of a $k$-essence model into a kinetic part and a potential part. We then derive the fluid equations for the kinetic density and the 4-velocity, with the EoS $P_k=\rho_k$, and complement them with the evolution equation for the scalar field to compute the value of the spatially dependent quantities, like $c_s^2(\phi, X)$. Different from the equations in previous works, the resulting equations, Eqs.~\eqref{continuity2: k-essence}--~\eqref{phi2: k-essence}, are fully nonlinear and accurate, in the sense that their derivation does not need small-quantity expansions of the density field and the estimation of the EoS. They are also fully self-consistent, in the sense that the number of equations equals the number of unknown variables, and everything is derived from a Lagrangian.

This action-splitting method offers several additional advantages. First, only discretizing the kinetic energy and making the potential play the role of a source, which transports additional energy to the kinetic part, is consistent with physical intuition. Furthermore, though deriving the EoS at nonlinear scales is not impossible, evolving the kinetic part of DE in the fluid equations appreciably simplifies the EoS to $P_k=\rho_k$, which holds at all scales. This will render the implementation of the fluid method more straightforward and efficient.

As a preliminary application of the framework, we perform spherical-collapse simulations involving both DM and DE components for two representative $k$-essence models, referred to as the XE and SV models. The simulations follow the development of DE density perturbations during the gravitational collapse of DM halos, illustrating the evolution of their radial density profiles at different redshifts. For the fiducial setup with $\bar{c}_s^2=10^{-6}$ and $M_{\rm halo}=10^{14}M_\odot$, the central DE density contrast reaches order unity by $z=0$, demonstrating the ability of our method to follow DE clustering into the nonlinear regime.

We further investigate the dependence of DE clustering on the background sound speed squared and the DM halo mass. The results show that lower sound speeds generally lead to stronger DE clustering due to reduced pressure support, while more massive halos produce density perturbations extending over larger spatial scales. In the limit of very small sound speeds, the DE density profiles develop nearly uniform central regions with relatively sharp transitions near the halo boundaries. These simulations illustrate the applicability of our framework under different physical conditions.

Throughout this paper, we focused on the XE and SV models as illustrative examples of clustering $k$-essence models, but the fluid framework developed here is not restricted to these particular theories. In principle, it can be applied to a much broader class of $k$-essence models, provided that the corresponding Lagrangian 
is specified. This flexibility allows the nonlinear evolution of clustering DE to be investigated within a variety of underlying scalar-field theories.

Although the present work is restricted to spherical-collapse calculations, our method can in principle be extended to fully three-dimensional simulations. Since the equations derived here are formulated in a Lagrangian description, they provide a natural starting point for implementation in particle-based numerical schemes, such as smoothed particle hydrodynamics (SPH) \cite{SPH01}. In particular, this framework provides a foundation for incorporating clustering dark energy into existing cosmological hydrodynamics codes, such as SWIFT \cite{SWIFT01} and GADGET-4 \cite{GADGET401}, thereby enabling future investigations of nonlinear DE evolution in fully three-dimensional cosmological simulations. The development of a full clustering DE SPH simulation code is currently underway.


In conclusion, our fluid framework offers a practical approach for incorporating the nonlinear evolution of dark energy into cosmological simulations and for testing the viability of different dark-energy models. This, in turn, can improve our understanding of how dark-energy dynamics affect cosmological observables, such as the large-scale matter clustering, cluster abundance and dynamics, and weak-lensing signals. In a companion paper \cite{CompanionPRL}, we will discuss some of the physical implications of the spherical collapse simulations, including the potential to distinguish between different clustering DE models using their small-scale clustering signals.

\begin{acknowledgements}
YG and JW are supported by the National Key R\&D Program of China (No. 2022YFA1602901). YG acknowledges support from the China Scholarship Council (No.~202510740017). CMB and BL acknowledge support from the UK Science and Technology Facilities Council (STFC) Grant No.~ST/X001075/1. YG and BL are supported by the ERC Advanced Grant UNCA through the UKRI Frontiers Research Guarantee (No.~EP/Z533877/1). JW acknowledges support from the NSFC Grant (No.~12125302), and the CAS Project for Young Scientists in Basic Research Grant (No.~YSBR-062).
\end{acknowledgements}

\bibliographystyle{apsrev4-2}
\bibliography{reference}

\appendix

\section{Derivation of 
spherical collapse equations under the 
Newtonian gauge}
\label{app: fluid and poisson}

For completeness, in this appendix, we give a detailed derivation of the relevant equations listed in Section~\ref{SC equations} under the 
Newtonian gauge
\begin{equation}
    \text{d}s^2=-\left(1+2\Psi\right)\text{d}t^2+a^2\left(1-2\Phi\right)\delta_{ij}\text{d}x^i\text{d}x^j.
\end{equation}
Part of our derivation is done with the help of the Mathematica package xPand \cite{xPand}. 

In spherical collapse simulations, the two gravitational potentials, $\Psi$ and $\Phi$, are of order $\mathcal{O}(10^{-6})$, and the radial (comoving) peculiar velocity, $v_{r,c}$, is of order $\mathcal{O}(10^{-4})\sim \mathcal{O}(10^{-5})$. 
On the other hand, the kinetic energy density (as well as the pressure) can have large, $\mathcal{O}(1)$, perturbations at low redshift. 
In the following derivation for the continuity, Euler, and Poisson equations, we shall neglect all of the second-order terms, such as $v^2$ and $v^i\partial_i\Phi$.

We can expand the continuity equation~\eqref{continuity2: k-essence} and the Euler equation~\eqref{Euler2: k-essence} term by term under the 
Newtonian gauge. The 4-velocity in this gauge is
\begin{equation}
    u^a=\left(1-\Psi, v^i_c\right).
\end{equation}
It follows that, after some calculation, the continuity equation~\eqref{continuity2: k-essence} is
\begin{equation}\label{comoving continuity}
\begin{aligned}
    &\frac{\textrm{d}\rho_k}{\textrm{d}t}+\rho_k\left(1+c_s^2\right)\left[\partial_iv^i_c+3H\textcolor{gray!60}{-\,3\dot{\Phi}}\right]\\
    =&\;\left(1\textcolor{gray!60}{\,+\,\Psi}\right)\left[\frac{1+c_s^2}{2}\sqrt{-X}K_\phi + c_s^2\left(-X\right)^{3/2}K_{X\phi}\right],
\end{aligned}
\end{equation}
and the Euler equation (Regarding Eq.~\eqref{Euler2: k-essence}, $i$-component $-$ $0$-component $\times \,v_c^i$) is
\begin{equation}\label{comoving euler}
\begin{aligned}
    &\rho_k\left(1+c_s^2\right)\left[\frac{\textrm{d}v^i_c}{\textrm{d}t}+2Hv^i_c+\frac{1}{a^2}\partial_i\Psi\right]\\
    =&\textcolor{gray!60}{\,-\,c_s^2v^i_c\partial_t\rho_k}-\frac{c_s^2}{a^2}\left(1\textcolor{gray!60}{\,+\,2\Phi}\right)\delta^{ij}\partial_j\rho_k.
\end{aligned}
\end{equation}
In the above equations, we have already used the EoS $\rho_k=P_k$. 
We note that, if we substitute the definition of the kinetic density and the velocity into the above two equations,  it can be proved that Eq.~\eqref{comoving continuity} is equivalent to the scalar equation of motion of the $k$-essence model, while Eq.~\eqref{comoving euler} is simply an identity at linear order. 

We can also expand the Einstein equation
\begin{equation}
    G^\mu_{\;\;\nu}=8\pi GT^\mu_{\;\;\nu},
\end{equation}
to get the Poisson equation. The $00$-component is:
\begin{equation}\label{comoving Poisson 1}
    3H^2(1\textcolor{gray!60}{\,-\,2\Psi})\textcolor{gray!60}{\,-\,6H\dot{\Phi}}+\frac{2}{a^2}\nabla^2\Phi=8\pi G\rho_{\rm tot},
\end{equation}
and the trace of the $ij$-component is
\begin{equation}\label{comoving Poisson 2}
\begin{aligned}
    -\left(9H^2+6\dot{H}\right)\left(1\textcolor{gray!60}{\,-\,2\Psi}\right)\textcolor{gray!60}{\,+\,18H\dot{\Phi}+6H\dot{\Psi}\,}&\\\textcolor{gray!60}{+6\ddot{\Phi}}+\frac{2}{a^2}\nabla^2\left(\Psi-\Phi\right)&=24\pi G P_{\rm tot}.
\end{aligned}
\end{equation}

It is also worth noting that several terms in the above equations, highlighted in gray, can be safely neglected in our spherical-collapse calculation. The reason is that the characteristic scale of the simulation considered here is sufficiently small that the dominant contributions associated with the gravitational potentials arise from the spatial gradients of the gravitational field in the Euler equation and from the Laplacian terms in the Poisson equation. The remaining terms involving $\Psi$ and $\Phi$ can become relevant when considering the large-scale and linear evolution of cosmological perturbations, but they have a negligible impact on the nonlinear spherical-collapse dynamics considered here.

In fact, we have explicitly included the two terms, $\Psi$ and $\Phi$, in the spherical collapse simulations. The resulting relative differences in the density profiles are found to be smaller than $10^{-4}$ throughout the evolution. In addition, we can compute the term $c_s^2v^i_c\partial_t\rho_k$ with the help of the continuity equation, and we checked that it also leads to a relative difference smaller than $10^{-4}$ in the density profiles, at least when $\bar{c}_s^2=10^{-6}$. The time derivative terms, like $\dot{\Phi}$ and $\ddot{\Phi}$, are slightly more complicated to implement, but we can still estimate their order of magnitude in our simulation. For instance, the gravitational potential at the edge of the DM halo is approximately
\begin{equation}
    |\Phi|\sim\frac{GM_{\rm halo}}{R_{\rm halo}c^2},
\end{equation}
and therefore, its time derivative is approximately
\begin{equation}
    H_0^{-1}|\dot{\Phi}|\sim\frac{GM_{\rm halo}}{R_{\rm halo}c^2}\left|\frac{\dot{R}_{\rm halo}}{H_0R_{\rm halo}}\right|\sim\Phi\left|\frac{\dot{R}_{\rm halo}}{H_0R_{\rm halo}}\right|.
\end{equation}
At redshift 0, $|\dot{R}_{\rm halo}|\sim0.1\;{\rm kpc/Myr}$, $R_{\rm halo}\sim1\;{\rm Mpc}$, and $H_0\sim10^{-4}\;{\rm Myr^{-1}}$. As a result, we have $H_0^{-1}|\dot{\Phi}|\sim\Phi$, and if adding $\Phi$ to our simulation essentially will not lead to any difference, 
we expect the same for $\dot{\Phi}$. A similar analysis can be performed for the $\ddot{\Phi}$ term.

\begin{figure*}[t]
\includegraphics[width=\textwidth]{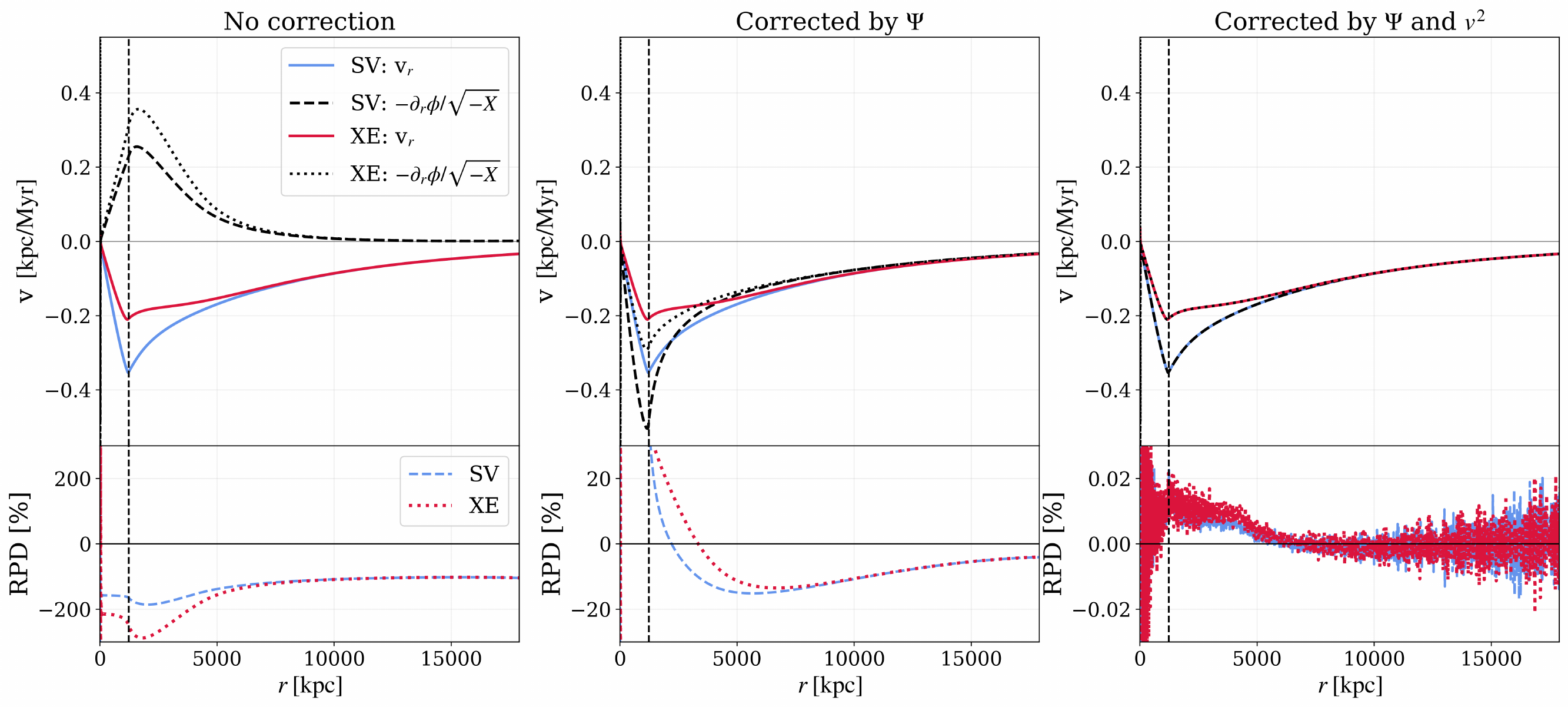}
\caption{Comparison between the radial velocity and the reconstructed $-\partial_r\phi/\sqrt{-X}$ at redshift 0 when using different evolution equations for $\phi$. {\textit{First column:}} The comparison when neglecting $\Psi$, $v_{r,c}^2$, and $\Phi$ in Eq.~\eqref{comoving phi}. The upper panel exhibits the velocity profiles (XE: red, SV: blue) as well as $-\partial_r\phi/\sqrt{-X}$ (XE: dotted line, SV: dashed line) versus the physical radius. The lower panel displays the relative percentage difference (RPD), $\left(-\partial_r\phi/\sqrt{-X}-v_r\right)/v_r$, for XE (red dotted line) and SV (blue dashed line) models. The vertical dashed line again marks the average radius of the DM halo within XE and SV models. {\textit{Second column:}} The same comparison when accounting for $\Psi$ in Eq.~\eqref{comoving phi}. {\textit{Third column:}} The same comparison when accounting for both $\Psi$ and $v_c$ in Eq.~\eqref{comoving phi}.
}
\label{fig:v_and_phi_gradient}
\end{figure*}

Finally, utilizing the relation
\begin{equation}
    \frac{\partial}{\partial t}\bigg|_{\rm comov}=\frac{\partial}{\partial t}\bigg|_{\rm phys}+Hr\frac{\partial}{\partial r},\quad\frac{1}{a}\frac{\partial}{\partial r_c}=\frac{\partial}{\partial r},
\end{equation}
with $r_c$ being the comoving radius, as well as $u_{\rm DE}=Hr+av_{r,c}$, $H^2=\dfrac{8\pi G}{3}\bar{\rho}$, and $\dot{H}=-4\pi G\left(\bar{\rho}+\bar{P}\right)$, Eqs.~\eqref{comoving continuity}, ~\eqref{comoving euler}, ~\eqref{comoving Poisson 1}, and ~\eqref{comoving Poisson 2} can be transformed into the physical version of equations~\eqref{SC continuity}, ~\eqref{SC Euler}, and~\eqref{SC Poisson}.

Next, we explore the trickiest part, the evolution equations for the scalar field. From Eq.~\eqref{phi2: k-essence}, the time derivative of $\phi$ reads
\begin{equation}\label{comoving phi}
\begin{aligned}
    \frac{\text{d}\phi}{\text{d}t}&=\frac{\text{d}\tau}{\text{d}t}\sqrt{-X}\\
    &=\sqrt{1+2\Psi-a^2v_{r,c}^2\left(1\textcolor{gray!60}{\,-\,2\Phi}\right)}\sqrt{-X}.
\end{aligned}
\end{equation}
To examine which terms are and are not negligible in the above equations, we consider the relation between the (physical) peculiar radial velocity of dark energy and the gradient of the scalar field:
\begin{equation}\label{eq:v_radial_consistency}
    v_{r}=av_{r,c}\approx-\frac{\partial_r\phi}{\sqrt{-X}},
\end{equation}
where we still neglect correction terms such as $v_r^2/c^2$ and $\Phi$ on both sides of the above formula. In spherical collapse simulations, we calculate the velocity of every concentric shell in each time step, so we know the velocity profile of the DE component at any time. 
Meanwhile, we also evolve the scalar field through Eq.~\eqref{phi2: k-essence} and compute the value of $X$ through $\rho_k=XK_X(\phi, X)$ at each time step, which means we can numerically reconstruct the quantity $-\partial_r\phi/\sqrt{-X}$ for every DE shell. We next compare the radial velocity and the $\phi$-gradient term at $z=0$ in the fiducial simulations as a consistency and correctness check of our numerical scheme: the two sides of Eq.~\eqref{eq:v_radial_consistency} must agree with each other.

The results of the comparison are exhibited in Fig.~\ref{fig:v_and_phi_gradient}. In the first column, we display the DE velocity profiles as well as the reconstructed $-\partial_r\phi/\sqrt{-X}$ when omitting $\Psi$, $v_{r,c}^2$, and $\Phi$ in Eq.~\eqref{comoving phi}. Although the peculiar velocities for XE and SV models are negative, as they should be for spherical collapse, the reconstructed $\phi$-gradient terms are actually positive: the two sides of Eq.~\eqref{eq:v_radial_consistency} do not match even qualitatively. In the second column, where we apply the (dimensionless) $\Psi$ in Eq.~\eqref{comoving phi}, whose order of magnitude is around $\mathcal{O}(10^{-6})$, and still neglect $v_{r,c}^2$ and $\Phi$, the profiles of the reconstructed $\phi$-gradient terms now present a similar radial trend compared to their radial velocity counterparts. Nevertheless, inside the radius of DM halos, the relative differences between the velocities and the $\phi$-gradient terms are still more than 20\%. On the other hand, in the third column, we further add the (dimensionless) $a^2v_{r,c}^2$ term into the scalar evolution. The velocity squared term is of the order $\mathcal{O}(10^{-9})$, which is 3 more orders of magnitude smaller than $\Psi$, yet still plays a significant role in the reconstruction of the $\phi$-gradient terms: after 
accounting for it in the $\phi$-evolution equation, Eq.~\eqref{comoving phi}, the relative differences between $v_r$ and $-\partial_r\phi/\sqrt{-X}$ are reduced to 0.02\%. Finally, the term $a^2v_{r,c}^2\Phi$ is 6 more orders of magnitude smaller than $a^2v_{r,c}^2$, which essentially does nothing to the final results of our simulation.

We note that the velocity and density profiles obtained using the three different treatments of the $\phi$-evolution equation, Eq.~\eqref{comoving phi}, shown in the three columns of Fig.~\ref{fig:v_and_phi_gradient}, are nearly indistinguishable for each DE model. We have further verified that, for both models, the local values of the source terms (terms on the RHS of the continuity equations), $K$, and even $c_s^2$, which are theory-dependent and thus can affect the simulation results, remain largely insensitive to the neglected gravitational-potential and velocity-squared corrections in the evolution equation for $\phi$, provided that we use the correct continuity and Euler equations. These results suggest that the corresponding corrections to the scalar-field evolution equation can be neglected without significantly affecting the fluid evolution (they are necessary only when we use the numerical solution of $\phi$ to reconstruct the radial velocity according to Eq.~\eqref{eq:v_radial_consistency}). This simplification may help reduce the computational complexity of future numerical implementations.


\section{Convergence Test of the Code for Spherical Collapse}\label{app: convergence}

\begin{figure*}[t]
\includegraphics[width=\textwidth]{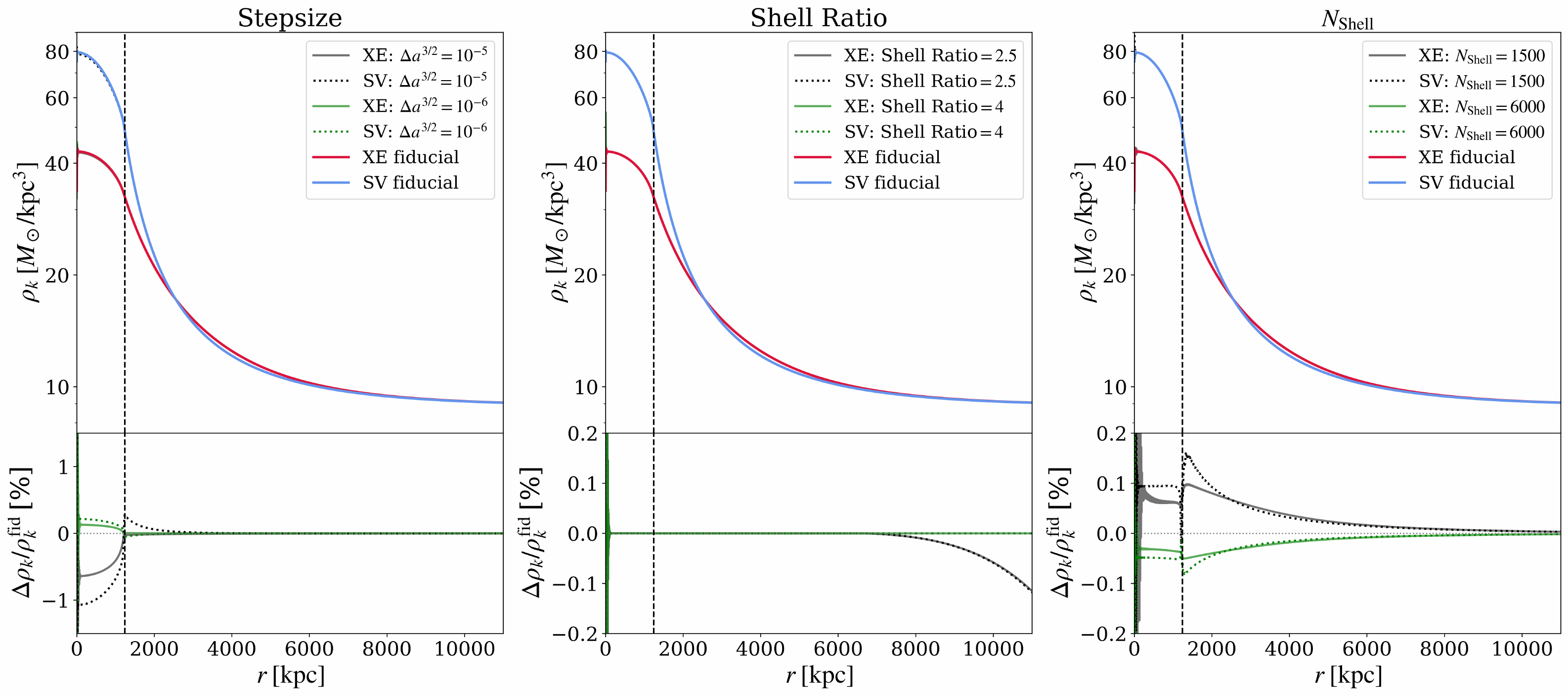}
\caption{The profiles for DE kinetic density in various convergence tests compared with the fiducial simulation (upper panel) and their relative differences (lower panel). The vertical dashed lines mark the average radius of the DM halo within XE and SV models. Each column shows a set of convergence tests in which a single numerical parameter is varied: from left to right, the integration time step, the ratio between the total number of shells and the number of shells that are inside the high-density region initially, and the total number of shells, respectively. See the text for a detailed description of these numerical parameters. Line styles and color schemes are indicated by the legends.}
\label{fig:convergence}
\end{figure*}

In this section, we test the convergence of our code by varying the numerical settings and quantifying the corresponding changes compared to the fiducial simulations. We mainly consider 3 numerical parameters:

\textit{Stepsize:} In our fiducial simulation, we use $\Delta a^{3/2}=2.5\times10^{-6}$ as the integration step size. In terms of convergence tests, we increase or decrease the step size to $\Delta a^{3/2}=1\times10^{-5}$ or $\Delta a^{3/2}=1\times10^{-6}$, for both XE and SV models, while keeping other parameters unchanged.

\textit{Shell Ratio:} This term denotes the ratio between the total number of shells and the number of shells that are inside the high-density region initially. To be specific, we have defined the initial condition in Eq.~\eqref{initial tophat} and initially we use 1000 shells inside $R_{\rm halo,ini}$ and another 2000 shells outside the boundary of the halo (or 3000 shells in total). We define the shell ratio to satisfy
\begin{equation*}
    N_{\rm shell}=({\rm shell\;ratio})\cdot N_{\rm shell}(<R_{\rm halo,ini}),
\end{equation*}
with $N_{\rm shell}=3000$ and $N_{\rm shell}(<R_{\rm halo,ini})=1000$ for both DM and DE. This means that in the fiducial simulation the shell ratio $=3$. With $N_{\rm shell}(<R_{\rm halo,ini})=1000$ and $R_{\rm halo,ini}$ unchanged, we also run simulations whose shell ratio $=2.5$ and $=4$ for XE and SV models, with other parameters unchanged. It is worth noting that the difference in the shell ratio only changes the radius of the outermost shell and does not affect the spatial resolution in the simulations (to test the effect of spatial resolution we have to tune $N_{\rm shell}$, which is described next).

$N_{\rm shell}$\textit{:} The total number of shells used in the simulation for DE or DM. In the fiducial simulation, $N_{\rm shell}=3000$. In the convergence tests, we coarsen or refine the spatial resolutions and set $N_{\rm shell}=1500$ or $N_{\rm shell}=6000$, while keeping the initial radius of the outermost shell $R_{\rm outermost, ini}$ unchanged.

The results are exhibited in Figure.~\ref{fig:convergence}. For the sake of simplicity and clarity, we only compare the influence of different numerical parameters on the kinetic energy, or there will be too many curves in one figure. 

It can be seen from the upper panels that all of the density profiles of the convergence tests nearly overlap with the fiducial density profiles. In addition, in the lower panels, most of the changes in the numerical parameters only lead to a change of $\sim 0.1\%$. The only exception is when $\Delta a^{3/2}=1\times10^{-5}$, which decrease the central density by $\sim 1\%$. This suggests the step size $\Delta a^{3/2}=1\times10^{-5}$ might be too large for high-density regions. We also note here that the density decline of the shell ratio $=2.5$ case at large radii in the second column is caused by boundary effects, as the smaller radial extent of these simulations brings the affected region closer inward. However, these boundary effects do not affect the central region, which is more important for our analysis of the clustering dark energy. 

These results indicate that the simulations are numerically converged, as varying the numerical settings leads to only minor changes in the results.

\end{document}